# A Shiny micromapST App

Randall Powers[1], Wendy L. Martinez[2]
[1] US Bureau of Labor Statistics, Powers.Randall@bls.gov
[2] US Census Bureau, wendy.l.martinez@census.gov

**Abstract**
The linked micromaps approach was originally developed as an improvement to choropleth maps for displaying statistical summaries connected with spatial areal units, such as countries, states, and counties. Two R packages to create linked micromaps were published in 2015. These are the micromap and micromapST packages. The latter was originally for data indexed to the 50 US states and DC, but the latest version accommodates arbitrary geographies. The micromapST package handles the formatting needed for linked micromaps and offers several options for statistical displays (scatterplots, boxplots, time series plots, and more). The micromapST package is very useful and takes care of most details of the layouts, but it can be problematic specifying the data frames needed to create the desired graphic. Furthermore, exploring data through visualization is easier, faster, and more intuitive using a graphical user interface. This is the motivation behind the R Shiny micromapST app. This paper will serve as a brief tutorial and introduction to micromapST and the Shiny app using real-world data and applications. In this paper, we provide background information on visualizing geographically indexed data and linked micromaps in Section 1. Section 2 discusses the data sets used in two illustrative examples. Sections 3 and 4 describe the application interface and show how it can create linked micromaps. The paper concludes with comments and future work.

## 1. Background

The goal of this work is to review a visualization approach provided in the micromapST package and to highlight the need to develop an R Shiny application based on the package, thereby creating a more user-friendly product.

When economists and other data professionals wish to construct visualizations of US state-level geographic data, they have many options to try. These include bar charts, tables, bubble charts (sometimes called proportional symbol charts), and choropleth maps. Choropleth maps are widely used for good reasons. They are intuitive for most people to understand, but there are some limitations. Our previous paper (Powers, et al., 2024) covered in detail the benefits and limitations of each of these methods.

Linked micromaps provide visual summaries of statistics, data, and geographic distributions in a coherent and intuitive manner. They provide accessible tools for exploring and disseminating statistical information connected to geographic regions, including complex distributional and multivariate patterns. In general, linked micromaps consist of several columns, with one column depicting a set of small maps and areal or polygonal units that are linked across rows to other columns showing statistics of interest. For background information on linked micromaps, see Carr and Pierson (1996), Carr and Pickle (2010), Payton, et al. (2015), and Pickle, et al. (2015).

The `micromapST` package was created as an easy-to-use interface for creating linked micromaps where the geographies of interest are the 50 states and Washington, DC. Such geographic data are typically of interest in US federal statistical agencies. The examples we examine here will be limited to US state data. The latest version of the `micromapST` package will handle arbitrary geographies with user-supplied boundary files. The initial version of the R Shiny application described in this paper does not handle non-US state data at this time.

While the `micromapST` package makes it easier to create a linked micromap, the use of the `micromapST` package requires a steep learning curve. For example, the data frame specifying the layout and contents of the linked micromaps graphic can be difficult and tedious to produce. This is demonstrated in Section 3 and Figure 5. The R Shiny application we developed is designed to eliminate much of the programming tasks. It provides the user with a graphical user interface to easily explore one's data and to rapidly create the desired graphic.

## 2. Examples

The focus of Powers, et al. (2024) was to show how a stakeholder might use micromaps to better inform their decisions and provided several illustrative examples. For this paper, we wish to take the extra step of describing how the challenges of using the `micromapST` package led to the creation of a more user-friendly micromap experience via an R Shiny application. We focus on two specific examples: (1) the effects of the COVID Pandemic on the employment levels of the Leisure and Hospitality industry for the Bureau of Labor Statistics (BLS) Quarterly Census of Employment and Wages (QCEW), and (2) the wage differences for Police and Sheriff Officers in the BLS Occupational and Employment and Wages Survey (OEWS). We describe these two data sources in this section

### 2.1 QCEW

The first example of linked micromaps uses data from the QCEW. The QCEW publishes state-level data on the BLS website, which is publicly available for download. The QCEW has quarterly state- and county-level data for numerous industries, for employment and wages, as well as ownership (private or government). The BLS website has an interactive tool that allows the user to choose an industry and what measure to display. The results can then be rendered in a choropleth map and summary table. However, the data and map are for one point in time. To tell a story with the data over time, the user must piece together multiple maps. Confusing things further, the sequence of maps has different legend color thresholds (see Figure 1) making it hard to compare measures over time

Powers, et al. (2024) demonstrated how a coherent story could be told using linked micromaps (see Figure 2). They showed how micromaps provide more information on the values and magnitude of changes in employment, as well as the state rank order of the desired variable. Additional interesting behaviors can be observed in Figure 2. For example, Arizona had some extreme swings in employment over the period, as seen in the time series column, and Washington, DC had the largest percentage change in employment. Linked micromaps give the user the ability to explore the data and to gain insights into how the pandemic affected employment across the nation and through time. In the next sections, we show how to replicate this linked micromaps graphic using the R Shiny app.

## 2.2 OEWS

The OEWS program produces employment and wage estimates annually for approximately 830 occupations. A partial table displaying OEWS variables for the Police and Sheriff Officers occupation is shown in Figure 3. One might want to explore wage differences for Police and Sheriff Officers across the nation, but it is difficult to understand relationships and trends from large tables like this. Data tables can contain a plethora of information, but one must read across multiple rows and columns and perform mental calculations to assess differences. This is another example where linked micromaps would be useful for understanding the data.

A linked micromaps graphic in Figure 4 examines possible wage differences for Police and Sheriff Officers in metropolitan and rural areas. Regions are sorted by the Average Hourly wage at the state level. The national mean wage is used for the reference value shown as a vertical line. The arrows run from the minimum hourly wage to the maximum hourly wage in metropolitan (MSA) and rural (BOS) areas within a state. A scatterplot shows a linear relationship between the average hourly wages of sheriffs and police officers in metropolitan areas vs. rural areas. Note that the colors used in this example are suitable for those with color vision deficiency as opposed to the default colors used in the `micromapST` package as seen in Figure 2.

# 3. R Shiny Application

While simpler to use in many aspects (Pickle, et al., 2015), there is still a learning curve to create linked micromaps with the `micromapST` package, which might discourage potential users. It is still challenging for the novice user to take advantage of the features, and the process of creating the graphic does not lend itself to rapidly and easily explore data sets. Certain inputs are required (such as number of data columns) and others are optional (such as axes labels or map titles) and must be programmed in a specific order. It is not readily apparent how to do this until you really dig into the documentation and many help examples.

See Figure 5 for an example code snippet used to produce Figure 2. The panelDesc data frame specifies the layout of the graphic, and one can see that it is hard to picture where labels go, what data variables will be displayed, etc. It is also important to note that the linked micromaps graphic is sent to a graphics device (usually a file such as pdf, png, etc.), not the development environment (IDE). In this example, it is sent to a .png file. So, a user must specify the layout, open a graphics device with a filename, create the linked micromaps, and close the graphics device. The resulting graphic is then opened with some other tool or application. Recall that the goal is exploratory data analysis (EDA). Being able to change things easily via a GUI and see results immediately without accessing some other tool would be beneficial.

As time went on, users of BLS and Census data wanted to apply linked micromaps in their work, and it became apparent that an interactive Shiny application integrating `micromapST` functionality was needed. An interactive application would allow users to enter the information for the desired graphic using text boxes and dropdown menus, and the coding would be done behind the scenes. Also, visualizing the resulting linked micromaps inside the application would make it a true EDA tool. The Shiny application is a user-friendly option to harness the package's tools and would make it available to more people.

The R Shiny application has three user tabs. The first is called the Welcome tab. This is the tab the application defaults to when the user opens the application. The goal of this tab is to introduce the user to the application, to describe the other tabs, and to present an example micromap as a template, which will help the user with the input process. The input process where the user specifies the components of the linked micromaps (data set, number of data columns, labels, etc.) is contained in the second tab, which is called the `MicromapST` tab. This tab contains numerous user inputs that allow the user to enter a map title, column labels, graphics choices, and more. The third tab is called the `MicromapST` Output tab. This tab produces a micromap graphic based on the user inputs and allows the user to save their results to a file. In the next section, we will walk through two examples that will highlight each of the tabs in detail.

## 4. Demonstrating the `MicromapST` App

To better explain the R Shiny application and how to use it, we will present two examples. The first uses the QCEW data, and the second uses the OEWS data. While these two examples will not demonstrate all the user input options, the hope is that they will provide a well-rounded introduction to the user process.

### 4.1 QCEW Example

The user opens the application by opening it in R Studio. The app opens to the `Welcome` tab. This tab gives a quick introduction to the application and displays an explanatory example image that serves as a guide for entering the inputs on the next tab.

The user then clicks on the `MicromapST` tab. This is the tab where the user enters the specific inputs for the linked micromaps graphic they wish to produce. When hovering over an input option, the user can see descriptions that help clarify the different options for the input (e.g., text box, dropdown menu, etc.).

Not all inputs are shown when the user first opens the `MicromapST` input tab. Only the area containing universal inputs for a linked micromaps graphic appears as seen in Figure 6. Some inputs have defaults or a pre-specified list and will appear pre-populated. First, the user would choose the data frame that contains their data (`Choose Dataset`). The dropdown will contain any data frame that is available in the global environment for the current R session. In the QCEW example, "`Temprates`" is the desired data frame.

Next, one chooses the number of additional columns, which are the columns where data or statistics are displayed. Recall that the linked micromaps output is divided into columns. There are always two by default, one being the map of the states and the second being the state labels in sorted order. By *number of additional columns*, we are referring to the number beyond the first two. These will be the glyph columns that give the statistical details the user is trying to convey. If the user is programming in the console or writing an R program like the ones that produced Figures 2 and 4, then the number of columns is only limited by how many will fit on the page or the area delineated by the graphic device (Pickle, et al., 2015). The `MicromapST` app has been limited to three additional columns for space and coding reasons, which should be sufficient for most applications. For the QCEW example, there were three additional glyph columns. When this dropdown is selected, extra user inputs appear, so users can enter the information needed for the desired graphic (Figure 7).

Other universal inputs for the overall linked micromaps graphic include map shading, which allows the user to choose what type of shading for the sub-regions (e.g., states) they desire. For the QCEW example, we used `maptail`, which highlights states cumulatively from the top and bottom maps toward the median state. Other options are available; see Pickle, et al., (2015) for details. Or, one can explore the different types of shading by choosing another option in the dropdown and seeing the results in the output tab.

The user then enters the overall title for the linked micromaps graphic in the `Enter Map Title` text boxes. The title can have up to two lines or each can be left blank. For the QCEW example, we use a rather long first line: "Effects of COVID: QCEW percentage Change in One Year Employment." For the second title line, we used "Leisure and Hospitality," as that was the specific job category we were looking at.

The user must choose which variable they want as the sort variable. This will order the state regions along with the data linked to them. For the QCEW example, we chose variable X1 in the data set. The user must also specify whether they want the data in descending or ascending order. We used descending, so the state with the highest value for variable X1 will be at the top.

Now the user can enter the specifications for the three data columns that just appeared in the Shiny application. First, they will enter the labels for the data columns. The first two labels will go above that column, and the third is displayed at the bottom of the data column. For scatterplots and time series plots, there will be a fourth label that goes on the vertical axis for each perceptual group; see Powers, et al. (2024).

The next step is to choose the type of plot (i.e., glyph) for that data column using the `Choose the Plot Type` dropdown menu. This dropdown input contains all the available glyphs for a linked micromaps graphic. Users can get an idea of what each glyph will look like, as an example icon is visible on each of the dropdown options. An example image displaying the chosen glyph is shown below the dropdown.

When the user chooses the desired glyph or plot type, more input options appear that correspond to that specific glyph. The first additional column series selected is time series. When creating a time series plot, the user much choose a time series array that must exist in the R environment. This array contains the data at the time points necessary for creating the time series plot; see Pickle, et al., (2015) for details on the structure for this array.

The second additional column is a dot chart. In the case of a dot chart, the user needs to indicate the variable used for the dots, which is X1 in this example. For some of the glyphs, there is an input prompting the user to enter a `refval`. This is an optional feature that allows for a vertical dashed green reference value to be placed on the map. For the QCEW example, we used "0" to emphasize the break between a negative and positive percentage change value.

Finally, the user repeats this process for Column 3. The plot type is `Arrow`. When the user chooses this option, they see that they must choose the beginning and ending values of the arrow. For our example it was X1 and X9, respectively.

Once the user has all their inputs filled in (see Figure 7), they can advance to the `MicromapST Output Tab`. Here, with the simple click of a button, they can produce the desired graphic. A key benefit of the app is that if the user does not like what they see, then they can easily go back and change the inputs. Whereas if they are doing it using the console, they must navigate to their hard drive, open the image file in another tool, make changes in the R console, and repeat until they have the desired graphic. Using the application, they can save the final linked micromaps graphic once they have a satisfactory version. A dropdown menu allows them to specify a file type, and a button saves it to the folder of their choice. See Figure 8 for a screenshot of the final linked micromaps graphic for this example.

### 4.2 OEWS Example

If the user is working within the same instance of the application, choosing a new data set in effect erases their inputs from any previous run. They start completely anew and will need to redo all inputs. The app also works such that if the number of additional columns is changed, all previous inputs specific to columns are erased. So, the user needs to make sure they want to change either of those two inputs. For the OEWS data, the input data frame is called "`PolData`."

In this example, there were four extra columns (Figure 4). Recall that the maximum one can have in the app is three. If the user needs all four (or more), then they will need to use the `micromapST` package functionality at the IDE command line. But if three columns are acceptable, the app again presents the easier option.

A screenshot displaying the inputs for three glyph columns using the OEWS data is shown in Figure 9 and Figure 10. For this example, the three glyphs are two columns with arrows and the third with a scatterplot. The first arrow graph was the `Hourly Wage Range` for Metropolitan Statistical Areas (MSA), and the second arrow graph was the `Hourly Wage Range` for Balance of State (BOS). The scatterplot allows the user to explore the relationship between the wages for the MSA and the BOS.

Here we can observe that the scatterplot is one of the glyphs that has an extra label entry for the Y-axis. From a GUI design standpoint, we want this user input to appear only when it is a requirement.

Again, once all inputs have been made, the user proceeds to the output tab and can produce the micromap; see Figure 11. If the output is what the user wants, then they save their work on their file directory. If it is not, then they go back to the input tab and adjust accordingly. This all occurs within the application, eliminating the need to open the image file as would be necessary if creating a linked micromaps graphic in the console using the functions.

## 5. Future Work and Final Comments

Recent additions to the application have included time series and box plot glyphs, these required additional input beyond what the other glyphs do. Additional help and color safe palettes are still to be added. Then the programming will be finalized, including error checking of the application. We will then write several vignettes illustrating the functionality of the application. We plan to submit the package to CRAN once it is ready for public release in 2026.

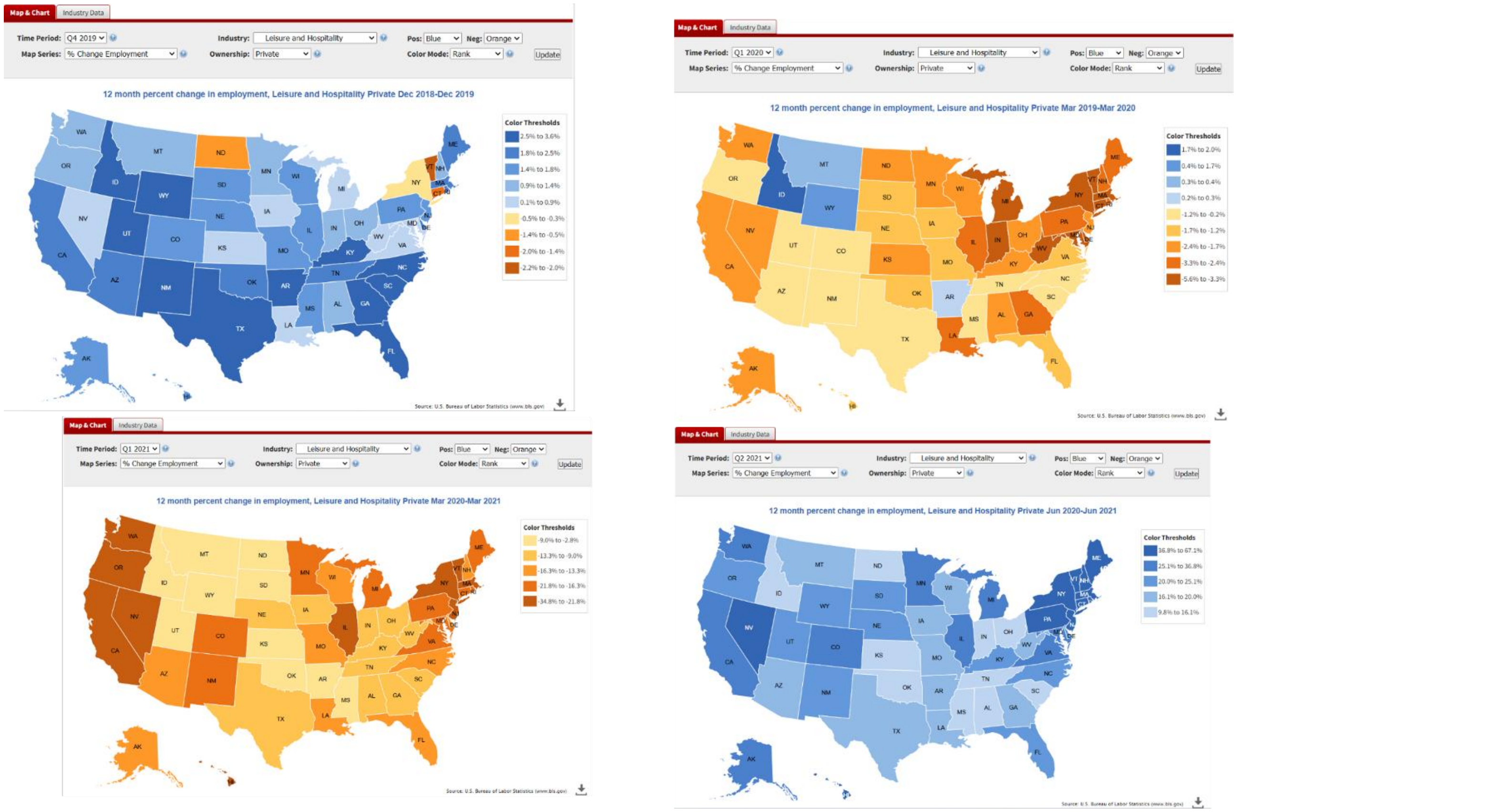

**Figure 1**. This figure depicts the Percent Change in Employment for four successive time periods for the Leisure and Hospitality Industry during the Covid-19 Pandemic. RETURN

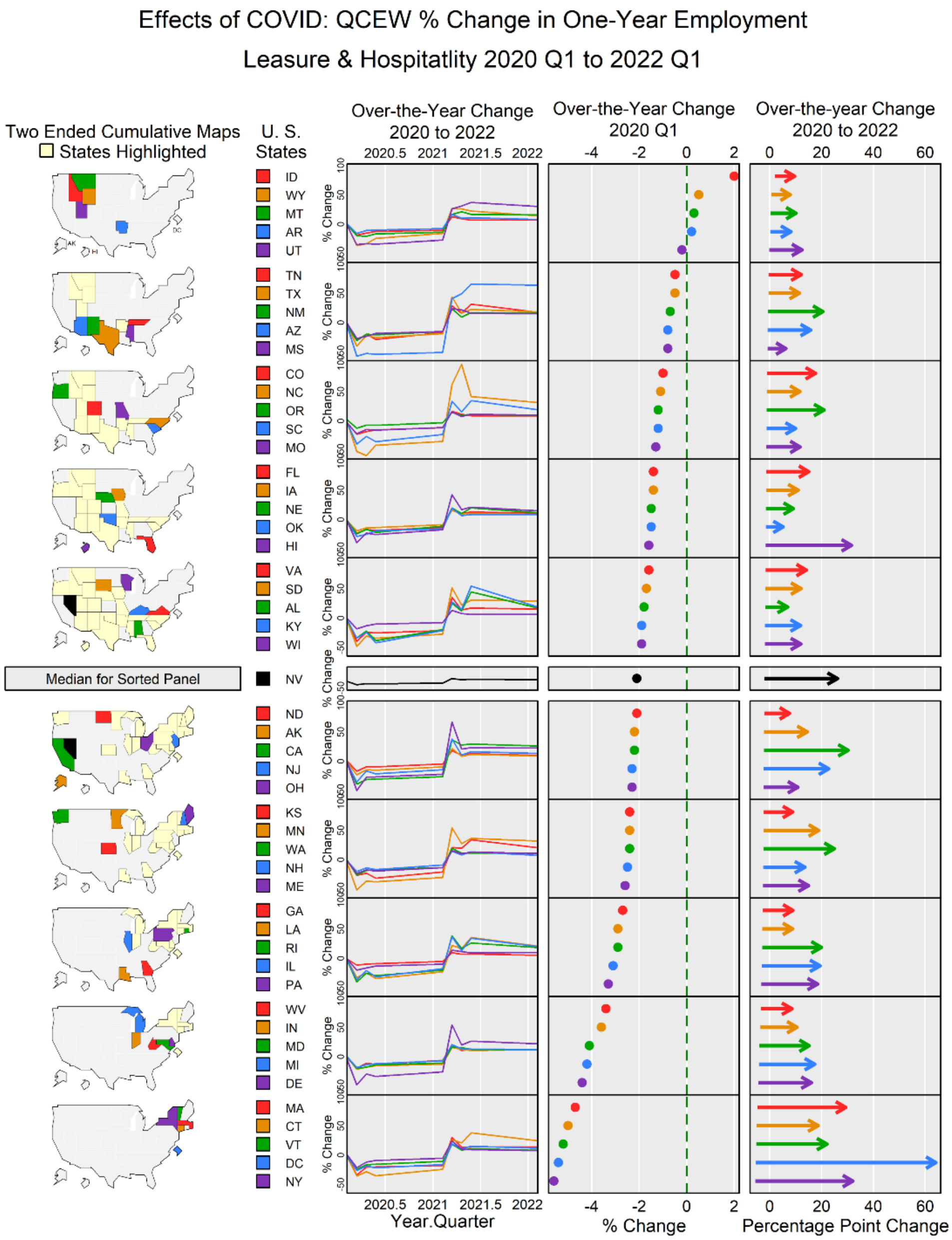

**Figure 2**. This figure depicts a micromap of the same data as Figure 1. Micromaps provided more information on the values and magnitude of changes, as well as the state rank order of the desired statistic. RETURN

| STATE | TOT_EMP | EMP % Err | JOBS/ 1000 | LOC_QUO | H_MEAN | A_MEAN | MEAN % Err | H_PCT10 | H_PCT25 | H_MEDIAN | H_PCT75 | H_PCT90 |
|---|---|---|---|---|---|---|---|---|---|---|---|---|
| AK | 1280 | 4.4 | 4.11 | 0.96 | 45.51 | 94660 | 2.2 | 27.88 | 36.8 | 43.3 | 52.98 | 64.3 |
| AL | 12770 | 1.8 | 6.22 | 1.46 | 25.13 | 52270 | 0.6 | 16.7 | 19.62 | 24.53 | 30.42 | 33.5 |
| AR | 5270 | 2.9 | 4.15 | 0.97 | 22.54 | 46880 | 0.8 | 15.99 | 18.27 | 21.3 | 25.94 | 33.78 |
| AZ | 12580 | 0.8 | 4.02 | 0.94 | 36.73 | 76390 | 0.4 | 26.89 | 31.51 | 36.76 | 41.98 | 44.65 |
| CA | 68010 | 0.6 | 3.79 | 0.89 | 53.74 | 111770 | 0.6 | 34.31 | 43.74 | 54.55 | 64.84 | 69.06 |
| CO | 9950 | 1.2 | 3.51 | 0.83 | 41.75 | 86840 | 0.6 | 29.84 | 36.29 | 42.5 | 48.81 | 50.89 |
| CT | 6660 | 1.2 | 4.01 | 0.94 | 39.04 | 81190 | 0.4 | 28.66 | 33.54 | 41.03 | 44.69 | 48.79 |
| DC | 5010 | 0 | 7.14 | 1.68 | 39.82 | 82820 | 0 | 28.94 | 34.18 | 38.48 | 44.55 | 52.64 |
| DE | 1730 | 2.3 | 3.69 | 0.87 | 39.51 | 82180 | 0.6 | 28.62 | 31.99 | 39.62 | 46.88 | 49.88 |
| FL | 48030 | 0.6 | 5.02 | 1.18 | 37.73 | 78480 | 0.5 | 22.67 | 25.71 | 32.81 | 43.31 | 53.46 |
| GA | 23370 | 5.4 | 4.91 | 1.15 | 27.02 | 56200 | 1.6 | 19.28 | 22.68 | 25.83 | 30.69 | 36.57 |
| HI | 2380 | 0 | 3.86 | 0.91 | 43.2 | 89850 | 1.4 | 35.5 | 41.41 | 42.42 | 42.7 | 53.51 |
| IA | 4920 | 1.7 | 3.18 | 0.75 | 33.91 | 70530 | 0.6 | 24.3 | 29.63 | 33.3 | 38.93 | 43.69 |
| ID | 2980 | 4.5 | 3.62 | 0.85 | 31.81 | 66170 | 1.5 | 23.23 | 24.93 | 30.34 | 36.71 | 43.05 |
| IL | 30550 | 2.5 | 5.08 | 1.19 | 42.13 | 87630 | 0.8 | 25 | 34.27 | 47.32 | 49.29 | 52.69 |
| IN | 12430 | 1.8 | 3.94 | 0.93 | 32.5 | 67590 | 0.5 | 24.24 | 28.29 | 31.89 | 37.46 | 40.07 |
| KS | 5840 | 2.7 | 4.14 | 0.97 | 27.38 | 56950 | 0.8 | 19 | 22.32 | 25.56 | 30.89 | 40.03 |
| KY | 7180 | 5.4 | 3.66 | 0.86 | 24.78 | 51540 | 1.9 | 17.14 | 20.04 | 24.27 | 28.57 | 33.41 |

**Figure 3**. This figure depicts a partial OEWS table for Police and Sheriff Employment. Most of this information is depicted on the linked micromaps graphic in Figure 4. RETURN

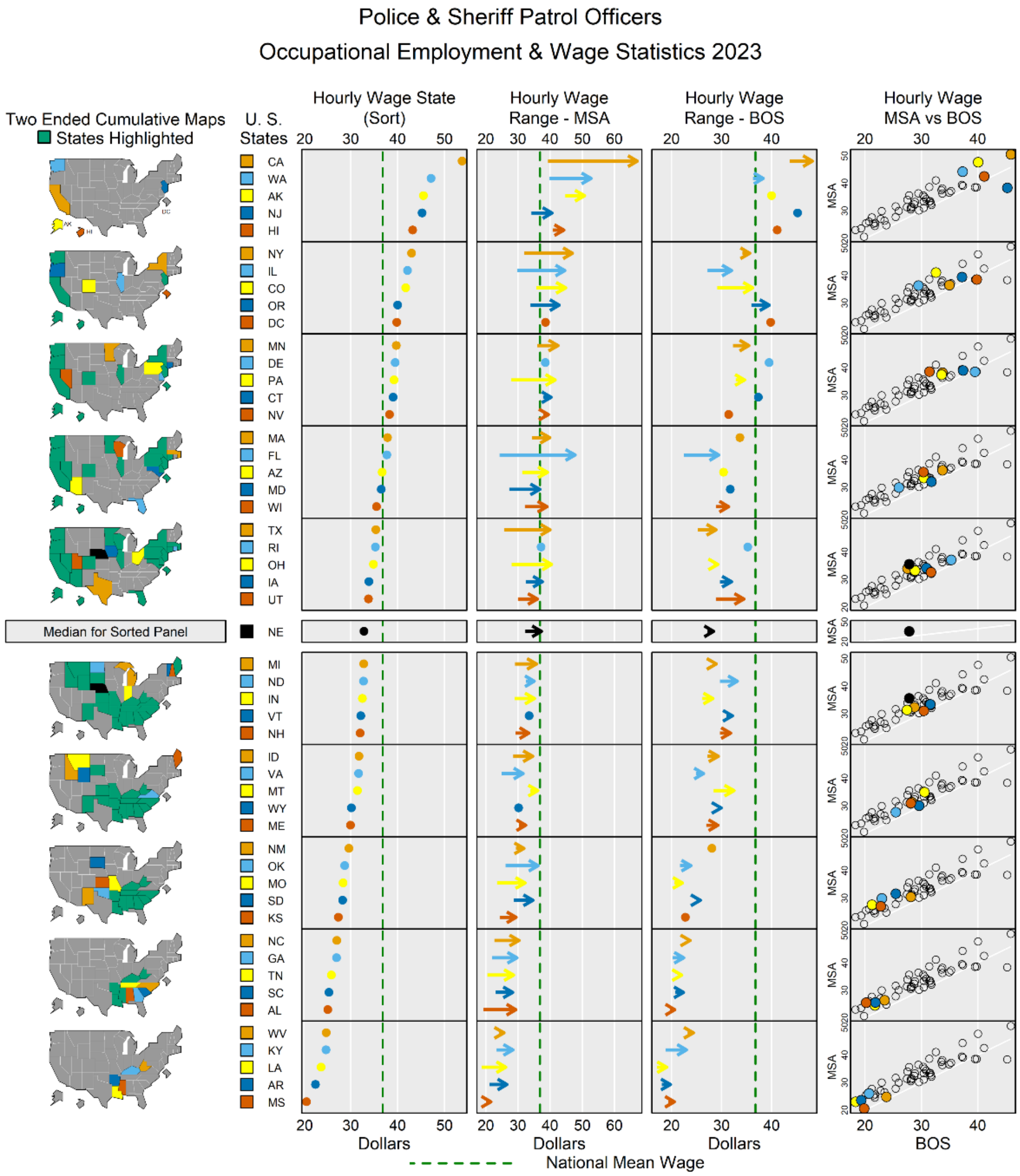

**Figure 4**. This figure examines possible wage differences for Police and Sheriff Officers in metropolitan and rural areas. RETURN

```
panelDesc <- data.frame(
  type=c("maptail","id","ts","dot","arrow"),
  lab1=c("","","Over-the-Year Change","Over-the-Year Change","Over-the-year Change"),
  lab2=c("","","2020 to 2022","2020 Q1","2020 to 2022"),
  lab3=c("", "","Year.Quarter","% Change","Percentage Point Change"),
  lab4=c("", "","% Change","",""),
  col1=c(NA,NA,NA,1,1),
  col2=c(NA,NA,NA,NA,9),
  refVals=c(NA,NA,NA,0,NA),
  panelData = c(NA, NA, "TSd", NA,NA)
)

ExTitle <- c("Effects of COVID: QCEW % Change in One-Year Employment",
             "Leasure & Hospitatlity 2020 Q1 to 2022 Q1")

grDevices::png(file="QCEWTimeSeriesArrow.png",
               width=7.5,height=10, units = "in", res = 600)

micromapST(temprates, panelDesc, sortVar=1, ascend=FALSE,
           title=ExTitle
)

x <- grDevices::dev.off()
```

**Figure 5.** This figure depicts the code used to produce Figure 2. Note the complicated format and required details for the panel description object that defines the layout of the linked micromaps graphic. RETURN

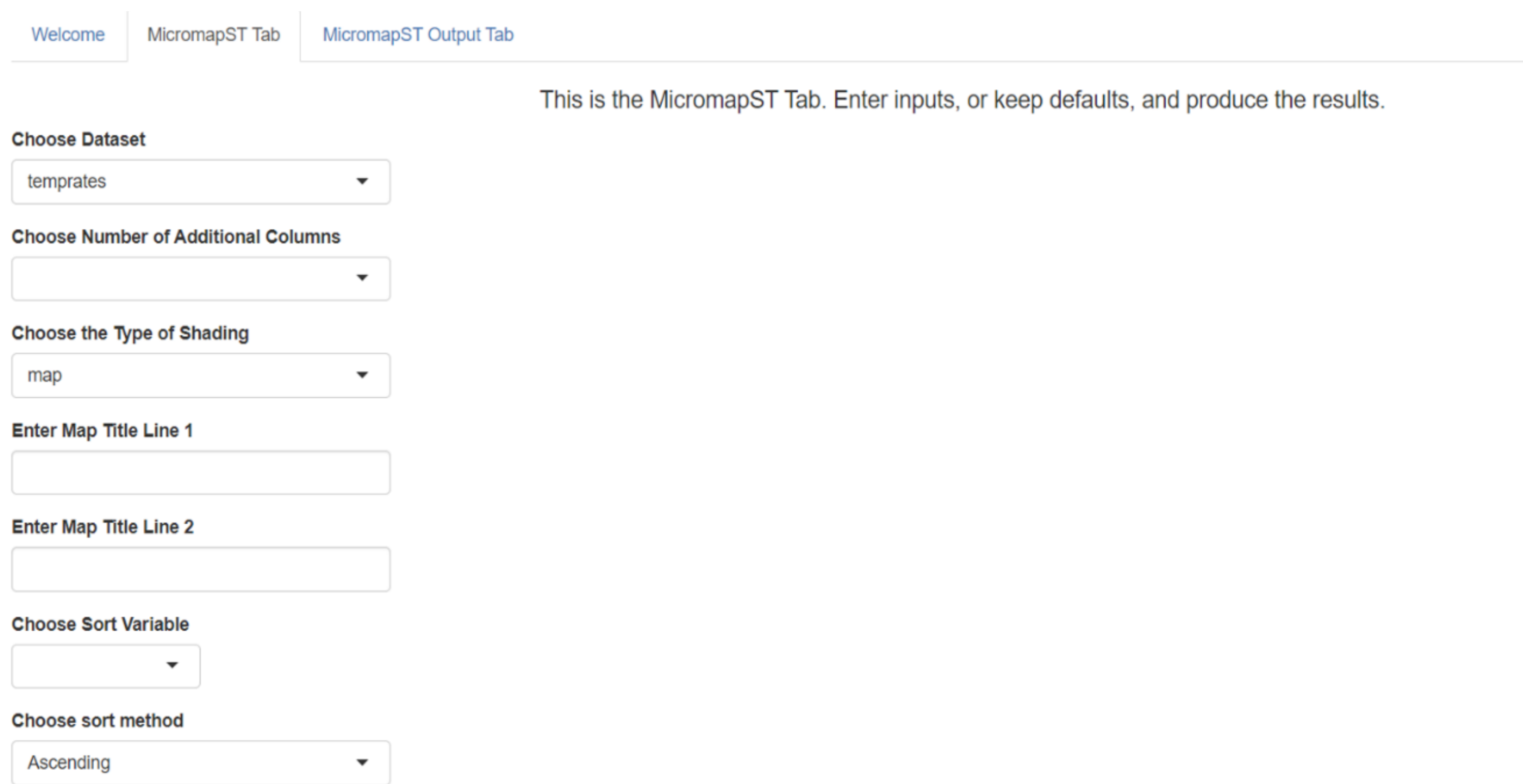

**Figure 6**. This figure depicts the `MicromapST` tab screen after the desired QCEW data frame is chosen, but before any other input is entered.
RETURN

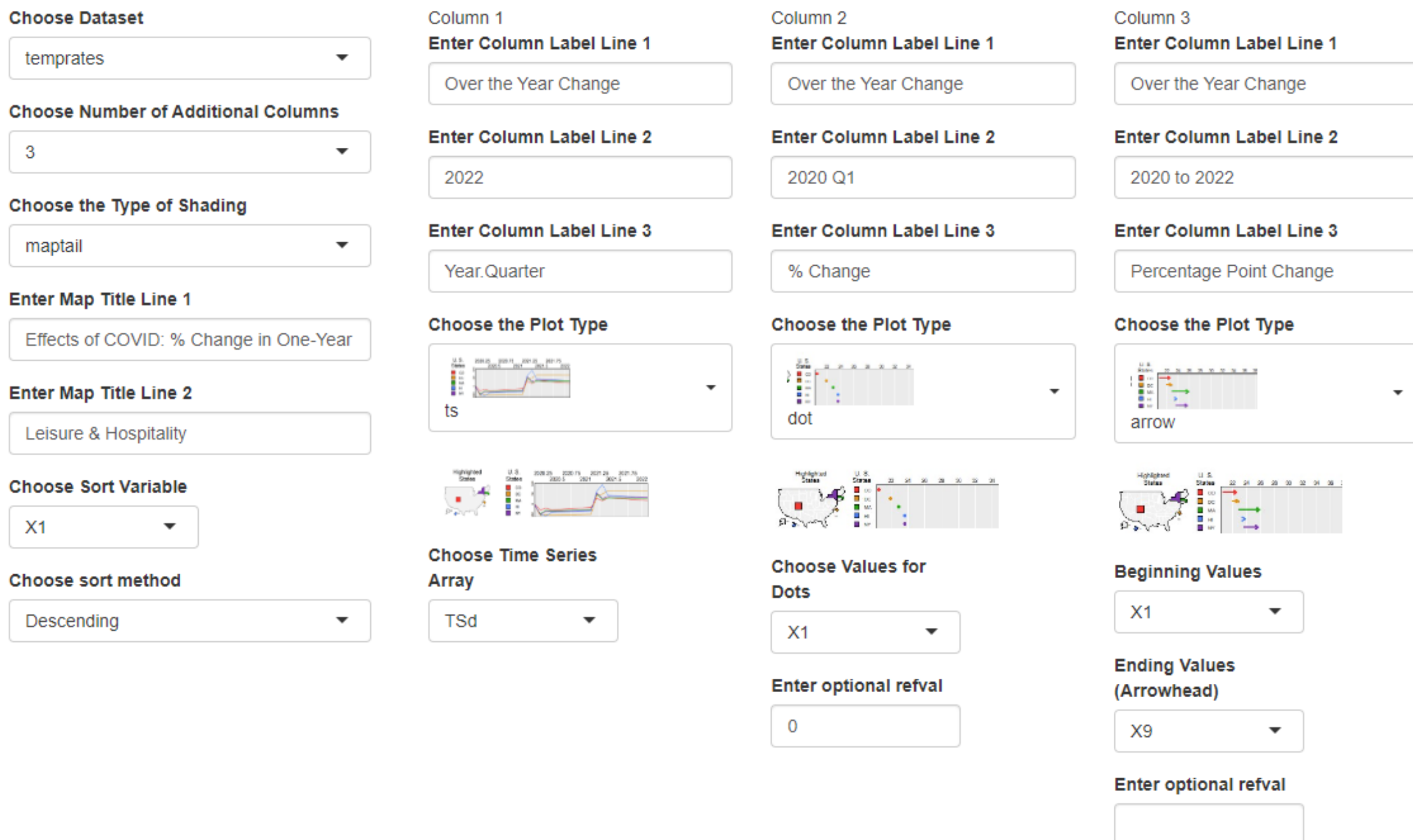

**Figure 7**. This figure depicts the `MicromapST` tab screen after all QCEW input has been entered. Since the user chose two additional columns, two additional column inputs columns appeared. RETURN

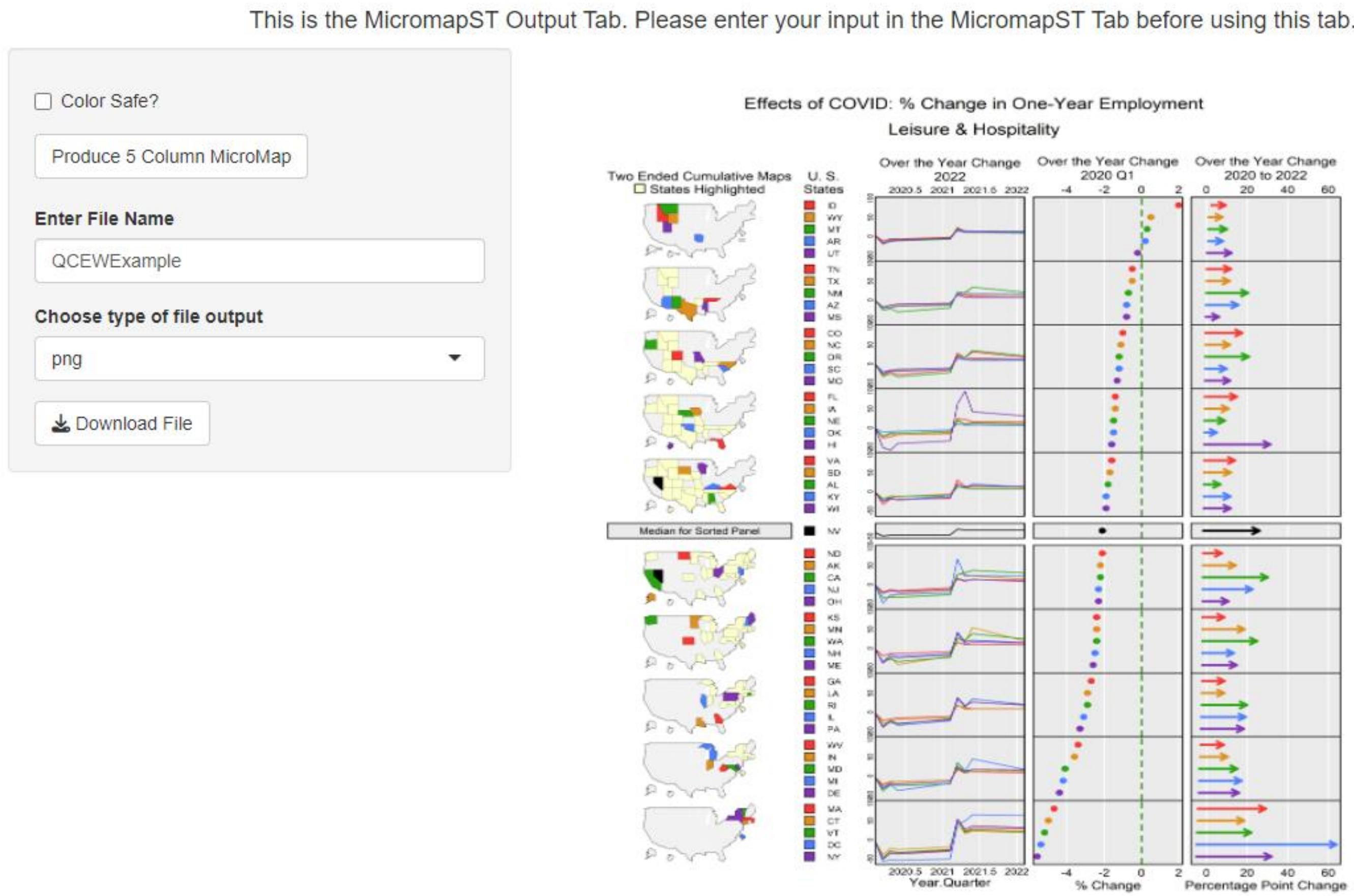

**Figure 8**. This figure depicts the QCEW demo output using the inputs from Figure 7. At this point, the user can choose to download the file. RETURN

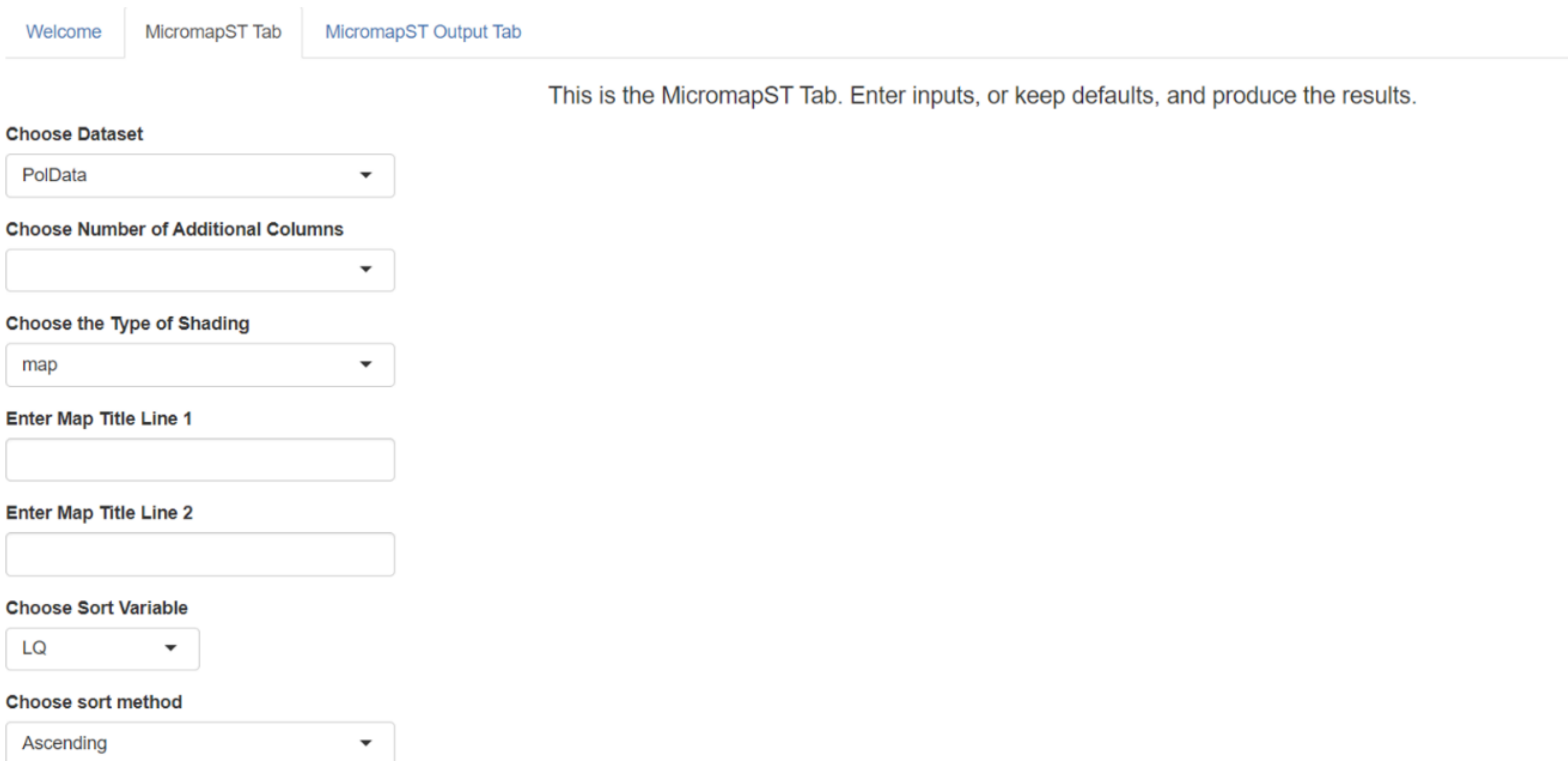

**Figure 9**. This figure depicts the `MicromapST` tab screen after the desired OEWS data frame is chosen, but before any other input is entered.
RETURN

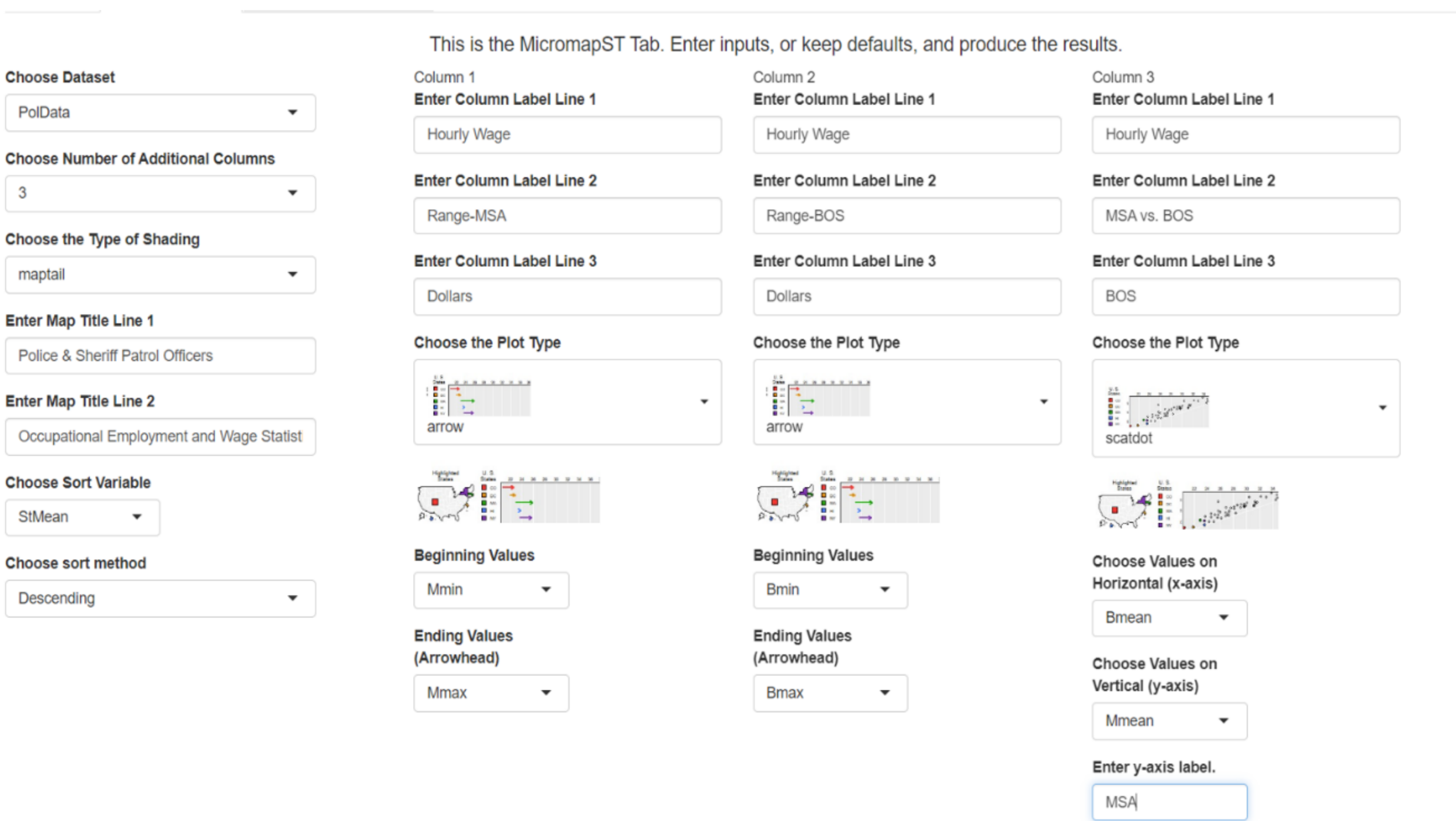

**Figure 10**. This figure depicts the `MicromapST` tab screen after all OEWS input has been entered. Since the user chose three additional columns, three additional column inputs columns appeared. Note the differing input prompts at the bottom of each column, depending on which glyph type the user chose. RETURN

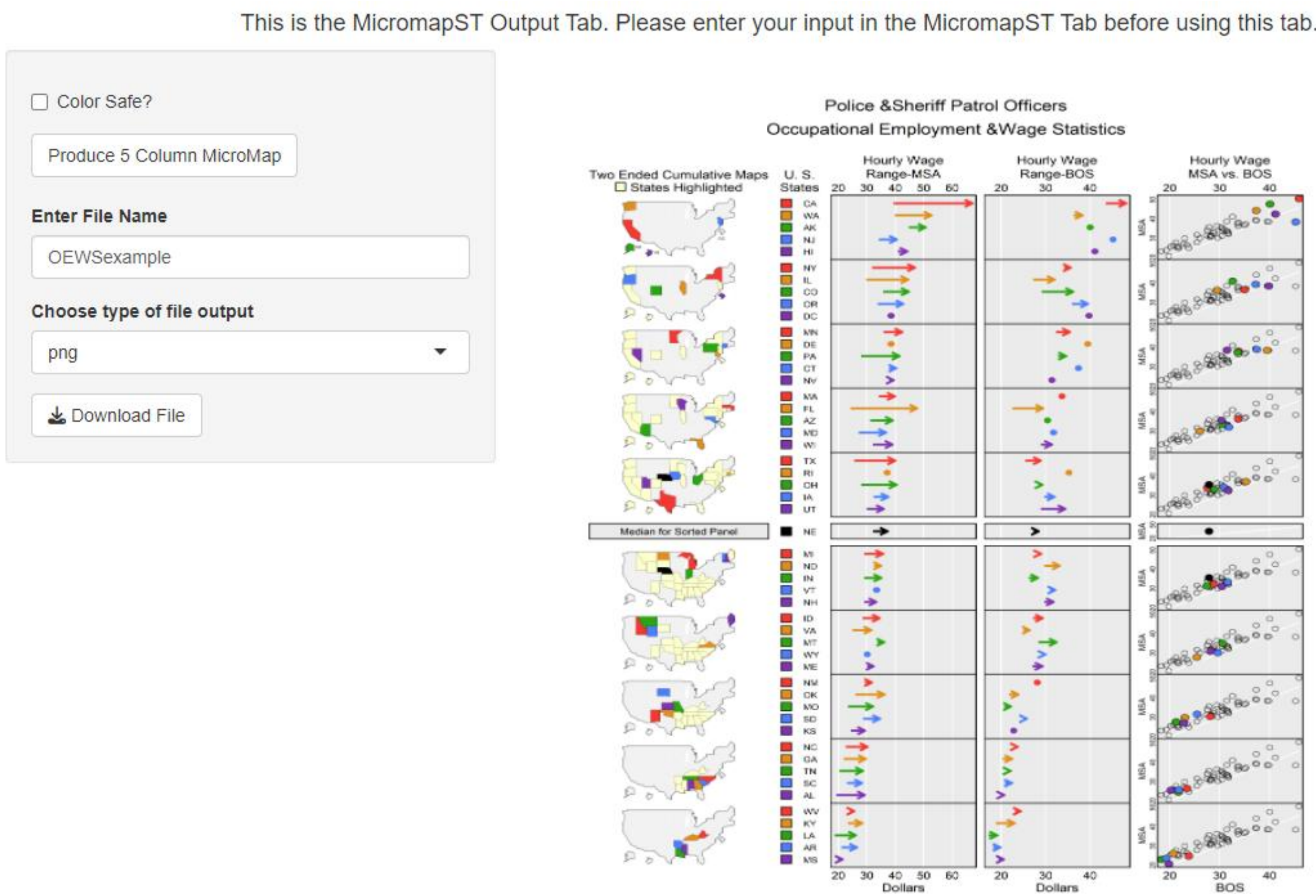

**Figure 11**. This figure depicts the OEWS demo output using the inputs from Figure 10. At this point, the user can choose to download the file.
RETURN